\documentclass[aps,prl,reprint,groupedaddress]{revtex4-2}

\usepackage{graphicx}
\usepackage{dcolumn}
\usepackage{bm}
\usepackage{hyperref}
\usepackage{multirow}
\usepackage{tabularx} 
\usepackage{float}

\begin{document}


\title{Creating a More Equitable Introductory Physics Classroom Through Invitational Phrasing in Question Solicitation}

\author{David Frykenberg}
\author{Brokk Toggerson}
 \email[Correspondence can be directed to ]{toggerson@physics.umass.edu}
\author{Adena Calden}
\author{Chris Ertl}
\affiliation{University of Massachusetts Amherst, Amherst, Massachusetts 01002, USA}

\date{\today}

\begin{abstract}
Asking questions during class time is one form of participation not commonly employed by members of underrepresented groups in large enrollment college classrooms, even though the form of participation is highly conducive to student learning. To encourage students to ask questions, many instructors initiate open question periods in their lectures with verbal solicitations (i.e., “Questions?”). Although several university centers for teaching and learning, such as at UC Berkeley, and instructor reflections suggest that the scripts of question solicitations impact the frequency of questions returned by students, no literature exists to support this claim. This quasi-experimental scholarship of teaching and learning (SoTL) study conducted with the Students-as-Partners model, observes the effect of a new solicitation script integrated into an introductory physics lectures to improve question participation from students. While the new solicitation script did not change the total number of questions asked by students in comparison to previous scripts, the number of questions asked by people who likely have gender identities traditionally underrepresented in physics (i.e. all but cis-men) increased significantly.
\end{abstract}

\maketitle


\section{\label{sec:Introduction} Introduction}

“Reflection is an integral part of [the Scholarship of Teaching and Learning (SoTL)] because it is an integral part of learning itself”~\cite{caf19}. Instructors who reflect on their teaching often note and lament the lack of questions from students. As student questions are an important form of classroom participation that largely benefits student learning, interventions made by instructors to increase student questioning are especially important in improving classroom equity, particularly in light of the noticeable discrepancy in the fraction of in-class questions asked by students from underrepresented groups. Thus, instructors clearly have a pivotal role in promoting equity in their classrooms through their influence on their students’ willingness to ask questions. 

Conventional wisdom suggests that an instructor can promote student questioning via the specific language used in question solicitation. For example, the Center for Teaching and Learning at UC Berkeley recommends using “What questions do you have?” as opposed to a more general “Questions?”~\cite{ctl}. Little, if any, SoTL research, however, exists to support such recommendations, and challenging such “conventional wisdom” is another key aspect of SoTL~\cite{n09}. 

This study aims to explore these issues: to determine if a change in how an instructor solicits student questions impacts participation with the intent of identifying ways to promote equity in learning. Following best SoTL practices, this study uses the Students-as-Partners model to provide diverse perspectives on the context of the course~\cite{f13}. Prior to leading this study, the lead author was a student in the course and then an undergraduate teaching assistant.

\subsection{\label{subsec:studentParticp} Student Participation and Questions in the College Classroom}

Higher education currently has a growing preference for active learning, participation-rich classrooms, as opposed to traditional lecture formats where students passively take in information~\cite{f12}~\cite{omh13}~\cite{w11}. The literature on active learning shares a consensus that in-class participation is beneficial to student learning~\cite{k19}~\cite{r10}~\cite{v95}~\cite{w11}. Students who participate in class are more academically motivated and exhibit higher levels of academic achievement in comparison to students who do not participate~\cite{k19}~\cite{omh13}~\cite{r10}~\cite{v95}~\cite{w03}~\cite{w11}. The common forms of in-class participation include: classroom discussion, small group involvement, as well as the willingness to ask and answer questions vis-à-vis the instructor~\cite{r10}~\cite{v00}. 

Among these forms, the simple practice of asking questions is especially important because of the wide range of learning benefits it provides students~\cite{a21}~\cite{a12}~\cite{co08}~\cite{hel03}~\cite{n21}. Students who pose questions to their instructor improve their problem-solving abilities and gain a more holistic understanding of material as it relates to other subjects~\cite{a21}. Those same inquisitive students who receive responses from their instructors become more motivated to learn in their courses~\cite{a12}~\cite{co08}~\cite{g12}. Furthermore, students who generate questions improve their study habits and show signs of reduced anxiety during examinations~\cite{a21}~\cite{hel03}. Peers of students who ask questions also benefit because they often harbor the same questions, gain new perspectives from questions they did not previously consider, and become more engaged in a class from the variety of questions added to the lesson-structure~\cite{n21}.

Unfortunately, it is likely that most students do not reap the benefits of asking questions in class or from general classroom participation. Colleges and universities commonly report a widespread lack of class-time student participation, specifically in the form of student questioning~\cite{co08}~\cite{hel03}~\cite{o13}~\cite{r03}~\cite{wq05}. Looking at the total number of questions asked in a lesson, teachers clearly ask more questions than students~\cite{a12}~\cite{hel03}. 

\subsection{\label{subsec:improvingequity} Improving Equitable Learning Through Classroom Participation}
 
In addition to classroom participation, equity in student learning is another focus of modern education. Equitable learning is defined by providing equal learning opportunities to students, regardless of gender or socioeconomic background~\cite{j20}. The motivation for integrating equitable learning into modern education is founded on the assertion that school classrooms are culturally biased environments which manifest as barriers relating to school organization, availability of resources, and classroom attitudes~\cite{c14}~\cite{shm19}. 

College classroom participation is one area where such barriers could potentially cause inequitable learning outcomes. Aside from the issue of its general lack in higher education, low classroom engagement is particularly pronounced in traditionally underrepresented social groups~\cite{et19}~\cite{n21}~\cite{w03}~\cite{w11}. Some reasons for this lack of engagement originate from the logistical structure of the classroom and from its method of instruction, including class size and the instructor’s communication style~\cite{b19}~\cite{k19}~\cite{r10}~\cite{wq05}~\cite{w03}. Other causes are derived from the student population of the class, such as student preparation, language barriers, and student emotions relating to participation~\cite{r10}~\cite{wq05}~\cite{w11}. 

One major factor contributing to the infrequency of student participation in college classrooms is stereotype threat. Underrepresented groups participate less than other students because they fear judgement and discrimination from their peers~\cite{et19}~\cite{w11}. Furthermore, when a student belongs to a traditionally underrepresented group, they may also avoid participating in class for fear of misrepresenting their culture~\cite{et19}~\cite{w11}. Coinciding with feelings of cultural alienation, underrepresented groups often do not engage in class because of their self-perception of intellectual and academic inadequacy~\cite{et19}.  

The goal of making classrooms more equitable and welcoming of student participation have yielded various teaching interventions. An instructor can generally raise in-class participation by setting standards for participation at the beginning of the course, acknowledging times for participation during a lesson, and giving students feedback when they participate~\cite{v00}~\cite{w03}. Additionally, instructors improve participation, especially with regards to underrepresented groups, by presenting themselves informally and by expressing openness to student opinion~\cite{g12}~\cite{retal10}~\cite{wq05}. The existing research on how to promote student questioning also shares the conclusion that student questions are primarily impacted by the instructor’s personable character and communication style~\cite{g12}~\cite{wq05}. One of the ways instructors communicate to their class and encourage questions in their students is by simply asking them if they have any questions~\cite{mc21}. However, as stated in the introduction, there is little SoTL research as to the efficacy of particular interventions such as the switch from “Questions?” to “What questions do you have?” which are commonly advocated by such institutions such as the UC Berkeley Center for Teaching and Learning~\cite{ctl}.

\subsection{\label{subsec:equity} Equity in the Physics Classroom}

Another of Felten’s Principles of Good Practice in SoTL is the grounding of the work within a particular context~\cite{f13}. In this case, the context is a large (~300 student) introductory physics course which is the second of a two-semester sequence for life science majors at a public R1 university in New England. 

If our goal is to promote equity in physics classrooms, then we must understand equity issues in physics. Within physics generally, students who do not identify as cis-men are traditionally underrepresented, and a great deal of literature explores the negative correlations between this minoritized status and performance~\cite{kpf09}. However, such overarching statistics may not be representative of the experiences of students who do not identify as cis-men in an introductory physics for life sciences course where the population generally self-identifies with she/her pronouns. However, the reflections of the authors as a former student/undergraduate TA and as the instructor of the course seem to indicate a disparity in questions asked between these two sub-populations. This perceived discrepancy is in line with the work of Eddy et al. who has also shown that the stereotype threat of women in science can extend to disciplines where the majority of students are not cis-men such as in biology~\cite{ebw14}. 

\section{\label{sec:method} Methodology}

The aim of the research is to gather data that would reveal whether there are common features of an instructor’s language in question solicitation that impact student participation. Two research questions guided this study:
\begin{enumerate}
\item{Are there common features of question solicitation which lead to an increase in student participation?}
\item{Is it possible to identify solicitation characteristics which promote a more equitable classroom through increasing the frequency of questions from traditionally underrepresented groups, in our case students who do not identify as cis-men?}
\end{enumerate}

Through analysis of recorded lectures, students' questions and presumed genders were tracked for every verbal solicitation initiated by the instructor of a physics classroom. Prior to the start of the study, the instructor of the course, following the advice of institutions such as CTL, UC Berkeley, had already chosen a solicitation script following the form, “what questions do you have?” for spring 2020. In contrast, during the spring 2019 semester, the instructor used whatever solicitation came to mind. However, the instructor consistently used variants of “Questions on this?” or “Any questions?” during this prior semester. 

\begin{figure}[tbh!]
	\includegraphics[width=\columnwidth]{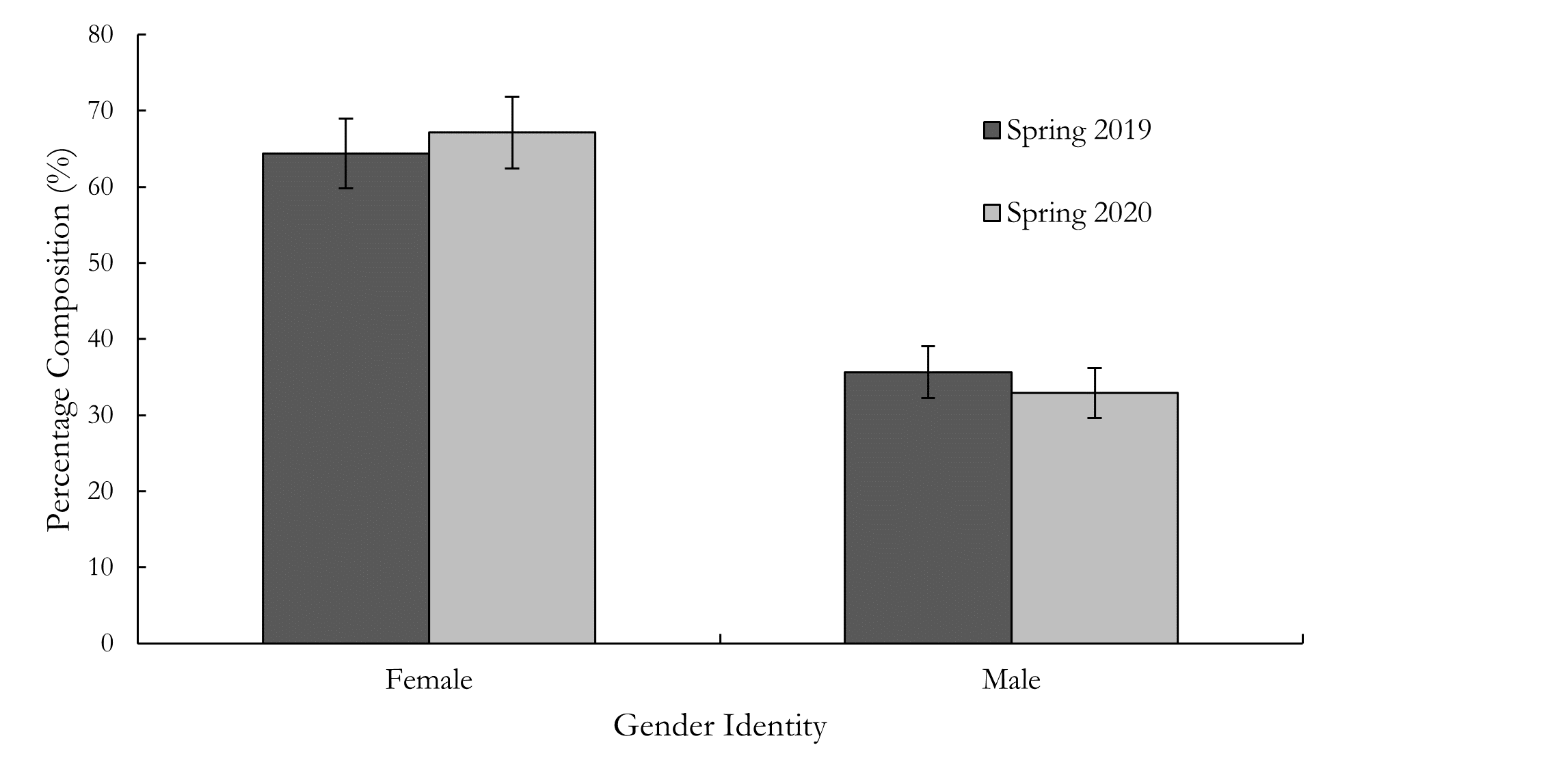}
	\caption{\label{fig:1} Distribution of student gender identities in 2019 and 2020. note: The percentage ratios, women to men, in 2019 and 2020 were 64.4\%:35.6\% and 67.1\%:32.9\% respectively. These data were taken from the beginning of each course before any students dropped out of the class. In each case, there was also a handful of students with other identities who do not statistically impact the outcomes of this study. }
\end{figure}

Data from this study uses the lecture recordings of a physics course that took place in the spring semesters of 2019 and 2020 at a large R1 university in New England. Specifically, the first two units of the course in each semester were used for observation as the standard teaching structure of the course changed in the latter half of 2020 due to the outbreak of SARS-CoV-2 which caused students to switch to remote learning. The class sizes from both semesters were approximately the same, with 303 students initially enrolled in 2019 and 301 students initially enrolled in 2020. Demographically disaggregated by gender, the 2019 class section had 64.4\% self-identify as women and 35.6\% self-identify as men. Meanwhile, the 2020 class section had 67.1\% self-identify as women and 32.9\% self-identify as men Figure~\ref{fig:1}. In each case, there was also a handful of students with other identities who do not statistically impact the outcomes of this study. 

Echo360, an automated lecture-capture system integrated into many STEM classrooms at the university, provided the recordings for data collection. For this course, Echo360 serves as an aid in making scientific demonstrations visible to the entire class and to enable students to review lectures including to those students who cannot attend a lecture because of extenuating circumstances. The class recordings provide visuals of lecture presentation slides as well as a video capture of the entire lecture hall from the back of the classroom. The audio comes from the instructor’s microphone.

Access to the Echo360 recordings was provided by the instructor of the course: a 38-year-old, white, cis-male who was born and raised in the US. Recordings were organized by lecture, course unit, and the year during which the Spring semester took place. For every recording, audio with corresponding lecture slides were observed for question solicitations made by the instructor.

The recording audio was the only source of question solicitation identification. The slides, however, provided context. Question solicitations were identified by two conditions: the instructor addressed the classroom’s student-body; and the instructor explicitly inquired into whether the students had any questions. A feature present in all solicitation scripts is the word, “questions,” a commonality which aided in the identification process. Only solicitations from the primary instructor, as opposed to one of the TAs circling the room, were considered and the time within the 50-minute lecture was also recorded. 

In addition, solicitations were also categorized as to whether they were initiated following the explanation of material or in response to a previous question posed by a student. Solicitations classified as “topic” (for “topic-initiated question solicitations”) were made after the instructor finished explaining some content. Alternatively, “sequence” solicitations (for “question solicitations initiated as part of a sequence of student questions”) come in response to a student-asked question during the class. A sequence was marked complete when no more questions were posed by students and the instructor moved on to other lecture content. A transition often also partially identified by the instructor, who might announce something along the lines of, ‘if there are no more questions, we will move on to…’. 

We classified any students who asked questions in response to the solicitation as cis-male or not based purely on listening to the recording audio (the students are not visible in the recording). Given that the significant majority of students who are not cisgender men identify as women, we use that term in our results though we acknowledge that there are trans-men and non-binary individuals in this group. The primary distinguishing characteristic is fundamental frequency based on linguists’ known practice of identifying an individual’s gender by his or her voice and the distinguishability of fundamental frequency amongst cis and trans individuals~\cite{lmb16}~\cite{hrwb20}. Prior to observation of the 2019 and 2020 data, researchers of the study conducted an inter-rater reliability test on data from Spring 2018 and found perfect agreement in their assessments.

\section{\label{sec:results} Results}

Students asked an equal forty-nine total questions in each semester. In the context of the first research question, this information implies that the 2020 solicitation script may not contain any new features over 2019 which significantly differ in their impact on the total number of questions asked by students. Alternatively, there may be new features present in the 2020 solicitation scripts whose effects are superseded by other factors not accounted for in the available data, such as the instructor’s time management of the lesson. 

In the spring 2019 semester, 38.8\% of the questions asked were from women (strictly, students not identified as cisgendered men). In spring 2020 by contrast, 67.3\% of the questions asked were from women. These percentages were calculated by dividing the number of questions asked by women in each semester by the respective total number of questions asked in that semester (Table~\ref{tbl:data}). 

\begin{table*}
\caption{\label{tbl:data} Solicitations Made, Total Number of Questions Asked, Questions Asked by Women, and the Percentage of Questions Asked by Women for Each Semester. Note: The percentages were calculated by dividing the number of questions asked by women in each semester by the respective total number of questions asked in that semester.}
\begin{tabular}{l l l l l}
\hline \hline
Semester & N Solicitations & N Questions & N Questions from Women & Average \%F \\
\hline 
2019 & 46 & 49 & 19 & 38.8\% \\
2020 & 43 & 49 & 33 & 67.3\% \\
\hline 
Difference & 3 & 0 & 14 & 28.5\% \\
\hline \hline
\end{tabular}
\end{table*}

A Pearson’s $\chi^2$-Test was used to measure the difference between the percentages in each semester. The binomial nature of the data and its independent collection in two separate semesters prompted the use of this test for statistical analysis. The null hypothesis specifies that the percentage of questions asked by women in the class section given the new solicitation script (spring 2020) does not differ significantly from the percentage of questions asked by women in the class section not given the new solicitation script (spring 2019). Conversely, the alternative hypothesis specifies that the percentage of questions asked by women in the class section given the new solicitation script does differ significantly from the percentage of questions asked by women in the class section not given the new solicitation script. As the number of questions asked in each semester was significantly greater than 5, the continuous $\chi^2$ distribution is a reasonable approximation for the discrete data in this study (Cochran’s Rule). 

The full results of the $\chi^2$ test are in Table~\ref{tbl:chi}. An $\alpha=0.01$ was chosen to assess the probability of falsely rejecting the null hypothesis from the $\chi^2$ test result. The data yield a 
$\chi^2 = 8.03$. As this $\chi^2$ test value exceeds 6.63, the $\chi^2$ value corresponding to $p=0.01$ with 1 degree of freedom, the null hypothesis is rejected. Instead, the alternative hypothesis, stating that the percentage of questions asked by women in 2020 does differ significantly from the percentage of questions asked by women in 2019, can be tentatively accepted. It is possible that the solicitations from 2020 are distinct from those of 2019 and contain a feature, or multiple, that is responsible for the observed increase in questions asked by women. 

\begin{table*}
\caption{\label{tbl:chi} $\chi^2$ Contingency Table and Test Result for Questions Asked by Men and Women in Each Semester. Note: For each cell, the numbers in parentheses indicate the expected values and the numbers in brackets indicate the test statistics.}
\begin{tabular}{l l l l l}
\hline \hline
Variable & F & M & Marginal Row Totals \\ 
\hline 
2019 & 19 (26) [1.88] & 30 (23) [2.13] & 49 \\
2020 & 33 (26) [1.88] & 16 (23) [2.13] & 49 \\
\hline
Marginal Column Totals & 52 & 46 & 98 (Grand Total) \\ 
\hline 
$\chi^2$ Test Result & \multicolumn{3}{l}{$\chi^2( \mathrm{dof} = 1, \; N = 98) = 8.03, \; \; p = 0.0046$} \\
\hline \hline
\end{tabular}
\end{table*}

\section{\label{sec:discussion} Discussion}

To determine what feature, or features, distinguished 2020 solicitation scripts from those of 2019, researchers of this study observed the data pools and found internal consistency amongst script formats within each of the semesters. All the solicitations from the 2020 data reflected the chosen style, “What…questions do [you all] have…?”. Likewise, with one exception, the 2019 solicitations followed two basic formats, “Questions on…this?”, and “Any…questions?” in order of frequency. The 2020 format contains an implicit assumption: that students already possess questions at the time of the solicitation. Conversely, the formats for 2019 simply probe the classroom to check that no questions exist on a topic before proceeding with the lecture. If students perceive such a perfunctory intent, it could be perceived as questions not being wanted. If these presumed interpretations match students’ actual experiences that would possibly help explain why female students in 2019 were less likely to ask questions than women in 2020. 
 
Although this study did not initially have this experimental design, the consistencies found in the solicitation scripts and within demographic data from 2019 and 2020 allow the results of the study to be more confidently interpreted as effects of solicitation formatting. Specifically, the consistencies provide support for the interpretation that women in 2020 asked more questions than women in 2019 because of the formatting of the 2020 solicitation script. Thus, although at its inception this study was observational, the consistencies in the data mentioned have allow this study to be redefined as quasi-experimental.

Such a shift in study design suggests some adjustments which should be addressed in future work. Although the consistency amongst the solicitations in 2019 allowed for comparison between 2019 and 2020, the instructor did not consciously choose a specific script in spring 2019 before the study began. Additionally, the while the instructor did choose the 2020 “What Questions do you have?” solicitation format to have the supposed implicit assumption that students already had questions, this choice was not backed by previous studies providing evidence on this solicitation format’s effectiveness on student participation. Moreover, there is no way to measure this supposed implicit assumption via an assessment tool or other means, because no such tool currently exists. Such a tool would facilitate a more thorough consideration of the particular features of various solicitations which may, or may not, promote equitable student questioning.

Even if such a tool to measure the implicit assumptions of different solicitations existed, determining the exact effect a solicitation might have on promoting questions from students would be challenging. Many factors contribute to the degree to which students are encouraged to ask questions. For example, students may feel more inclined to ask questions if someone else has already asked a question on the same topic. While student questions were categorized as “topic” or “sequence,” the results did not incorporate this categorization. Including this distinction may provide additional insights as the two categories may have distinct mechanisms of inspiring questions from students.

The technology recording the lectures’ audio also limited a deeper, more thorough analysis of the data. As a microphone on the instructor’s person was the sole source of the audio, students’ voices were audible, but the words of their responses were indiscernible. As such, distinguishing between student questions apart from other student vocalizations directly was impossible; only instructor comments allowed for an indirect method of differentiation. Moreover, the audio quality made identifying a single student who asked multiple questions impossible. Microphones situated throughout the room, perhaps in combination with video, could alleviate these concerns. 

\section{\label{sec:conclusions} Conclusions} 

This Students-as-Partners SoTL study finds that using a question solicitation script “What questions do you have?” in a second-semester introductory physics for life science students can produce a higher fraction of questions arising from students who are not cis-men, a group traditionally underrepresented in physics. We speculate that that the primary feature which results in this higher participation is an implicit assumption that students possess questions at the time of a solicitation. However, more research is required to confirm which features of specifically influence student questioning. More classroom data from various courses and instructors with various identities, would also strengthen the study, as would more thorough observation of lectures (via video or live observation) to better identify which students were asking questions. Ultimately, integrating measures of psychological features of classrooms and students may aid in developing an assessment tool for invitational properties of question solicitation. 

Aside from the central finding that women’s questions increased in response to the “What questions do you have?” question solicitation script, this study could not prove that this script had any new features over prior scripts affecting the total number of questions asked by students. Considering the results of this study, question solicitation claims made by centers for teaching and learning at UC Berkeley are partially supported, albeit with little specification for which features of question solicitation affect student participation. The results of this study hold promise for educational application and development of equitable learning environments. 

\section{Acknowledgments}

The authors wish to acknowledge the support of the University of Massachusetts Amherst for their financial support of this work through various internal grants with particular thanks to the Center for Teaching and Learning. We particularly thank Colleen Kuusinen for her help in developing this Students as Partners Scholarship of Teaching and Learning Survey. 

\bibliographystyle{apsrev4-2}
\bibliography{Invitational-Phrasing-arXiv}

\end{document}